\documentclass{aa}
\usepackage{graphicx}
\usepackage{amssymb,amsmath}
\usepackage{natbib}
\usepackage{color}
\usepackage{bm}

\def\msun{{\rm M}_\odot}

\begin{document}
\title{Compression of matter in the center of accreting neutron stars}
\author{M. Bejger\inst{1} \and J. L. Zdunik\inst{1}
\and P. Haensel\inst{1} \and M. Fortin\inst{1,2}}
\institute{N. Copernicus Astronomical Center, Polish
           Academy of Sciences, Bartycka 18, PL-00-716 Warszawa, Poland
\and
LUTh, UMR 8102 du CNRS, Observatoire de Paris, F-92195 Meudon Cedex, France\\
{\tt bejger@camk.edu.pl, jlz@camk.edu.pl,
haensel@camk.edu.pl, morgane.fortin@obspm.fr}}
\offprints{M. Bejger}
\date{Received 06/09/2011 Accepted 26/10/2011}
\abstract{}
{To estimate the feasibility of dense-matter phase transition, we
studied the evolution of the central density as well as  
the baryon chemical potential of accreting neutron stars. 
We compared the thin-disk accretion with and without the magnetic 
field torque with the spin-down scenario 
for a selection of recent equations of state.}
{We compared the prevalent (in the recycled-pulsar context) Keplerian
thin-disk model, in which the matter is accreted from the
marginally-stable circular orbit, with the recent magnetic-torque model
that takes into account the influence of stellar magnetic field on the
effective inner boundary of the disk. Calculations were performed using a
multi-domain spectral methods code in the framework of General Relativity.
We considered three equations of state consistent with the recently measured mass of PSR J1614-2230, $1.97\pm 0.04~{\rm M}_\odot$ (one of them softened by the appearance of hyperons).}
{If there is no magnetic torque and efficient angular momentum transfer
from the disk to the star, substantial central compression is limited to
the region of initial stellar masses close to the maximum mass.  Outside
the maximum mass vicinity, accretion-induced central compression is
significant only if the angular momentum transfer is inefficient.
Accounting for the magnetic field effectively decreases the efficiency of
angular momentum transfer and implies a significant central compression.}
{An efficient angular momentum transfer from a thin disk onto a
non-magnetized neutron star does not provide a good mechanism for the central compression and possible phase transition. Substantial central compression is possible for a broad range of masses of slowly-rotating initial configurations for magnetized neutron stars. Accretion-induced 
central compression is particularly strong for stiff equation of state 
with a high-density softening.}
\titlerunning{Evolution of central parameters in accreting NSs}
\authorrunning{Bejger et al.}
\maketitle
\section{Introduction}
\label{sect:intro}
Neutron stars (NSs) are detected in binary systems by X- and $\gamma$-ray
observatories because of the immense amounts of radiation released in the
process of accretion of matter falling from a companion star onto the NS
surface. Because the stellar mass increases during accretion, it is usually
assumed that the density in the center of the star increases as well, and that
a sufficient increase of mass may result in a phase transition (the
details of which are not yet fully known). It is expected that a phase
transition in the NS core might cause observable astrophysical phenomena, such
as star-quakes, spin clustering, gravitational wave emission and therefore
serve, it is hoped, as a testbed for the quantum chromodynamics (QCD) phase diagram (\citealt{ChengD1998,GlendenningW2001,BlaschkeGP2001,ZdunikHB2005,
BlaschkePG2008,XuL2009} and references therein).

We aim to show with three recent equations of state (EOSs) of NS
cores, that the central baryon number density, $n_{\rm c}$, and the baryon
chemical potential, $\mu_{\rm c}$,  may change in a non-trivial way while
the star gains in mass and simultaneously increases its rotation rate
during accretion. In particular, we will study whether, and if so, to
what extent, the evolution of $n_{\rm c}$ and $\mu_{\rm c}$  depends on
the EOS of dense matter. Our considered EOSs are consistent with the
recently-measured mass of PSR J1614-2230, $1.97\pm 0.04~{\rm M}_\odot$
\citep{Demorest2010}. Complementary questions that may be asked are (a)
how does the actual efficiency of the angular-momentum transfer and 
the stellar magnetic field coupling to the disk affect the final
rapidly-rotating NS configuration, and (b) how does the accretion-induced evolution depend on the mass of the initially slowly rotating configuration. Thus
far, the formation of millisecond pulsars was most often modeled following
the classical paper of \citet{CookST1994} who assumed the stellar
magnetic field is sufficiently weak that it does not influence the accretion flow significantly; consequently, one assumed $B= 0$ and the accretion
from the marginally-stable circular orbit (of radius $r_{\rm ms}$).
Notable exceptions were \citet{PossentiCGBD1999}, \citet{BurderiPCSD1999}, \citet{BlaschkeGP2001}, \citet{Colpi2001}, and \citet{BlaschkePG2008}, who used simple models of magnetic torque acting in the NS--accretion-disk system (\citealt{GhoshLamb1979}, see also \citealt
{GhoshLamb1991}).

Here we compare the results obtained using the $B=0$ approach with
those employing the recent effective magnetic-torque model of \citet
{KluzniakR2007}, used recently by \citet {BejgerFHZ2011} to study the
formation of a millisecond pulsar PSR 1903+0327. The details of
magnetic-field decay during the accretion-driven evolution phase of an 
NS' life are still far from being understood in spite of theoretical and
observational efforts in recent years (see Sect. 4 of \citealt
{BejgerFHZ2011} for a brief summary). Hence, we employ the simplest possible
description of the magnetic-field decay that was shown to be consistent
with at least a subclass of observed accreting NSs \citep
{ShibazakiMSN1989} - this approach allows us to study the qualitative features
of the influence of disk accretion on the behavior of the NS central parameters.

The text is organized in the following way. Sect.~\ref{sect:methods} gives
a short description of the methods used to obtain the accreting NS tracks. 
The results are presented in Sect.~\ref{sect:results}. Sect.~\ref
{sect:conclusion} contains our summary and conclusions.
\section{Methods and models}
\label{sect:methods}
We simulated the spin evolution of NSs caused by the thin-disk 
accretion without a magnetic field and compared it with the results 
for the magnetic-torque model of \citet{KluzniakR2007}. Because the 
original Newtonian magnetic-torque model was insufficient in extreme 
cases of rapidly-rotating, massive and compact NSs, we additionally 
employed the  modification by \citet{BejgerFHZ2011}, which takes 
into account the relativistic effects caused by the existence of the 
marginally-stable circular orbit (recalled briefly in the appendix). 
We considered three recent EOSs - APR EOS by \citet {AkmalPR1998}, 
DH EOS by \citet {DouchinH2001} and BM EOS from a set of non-linear 
relativistic mean field models of \citet{BednarekM2009}\footnote{We 
use a specific model corresponding to the parameter $\Lambda_{\rm 
V}=0.016$. It fits semi-empirical hypernuclear and nuclear data and 
yields for an NS rotating at $317$ Hz (spin frequency of PSR 
J1614-2230) $M_{\rm max}=2.01\ M_\odot$.}. The first two EOSs assume 
nucleon NS cores, while the BM EOS has a characteristic high-density 
softening associated with appearance of hyperons. Constructing 
an EOS yielding $M_{\rm max}>1.97\ M_\odot$ in spite of the hyperon 
softening is not an easy task, therefore it deserves some additional 
explanations. The crucial feature of our BM EOS is a repulsive 
contribution to the pressure coming from the quartic terms in the 
vector meson fields in the Lagrangian \citep{BednarekM2009}. We emphasize, that the BM model reproduces semi-empirical 
nuclear-matter data as well as the semi-empirical estimates of the 
potential wells of hyperons in the nuclear matter, coming from 
hypernuclear data, involving hypernuclei and $\Sigma^-$-atoms, as 
well as the strength of the $\Lambda-\Lambda$ attraction in the 
$\Lambda\Lambda$-hypernuclei \citep{BednarekM2009}. Neutron-star configurations were obtained with the numerical library {\tt LORENE}
\footnote{{\tt http://www.lorene.obspm.fr}}, using the 
implementation of the \citet{BonazzolaGSM1993} formalism for 
axi-symmetric and rigidly-rotating stars ({\tt rotstar} code).

\subsection{Magnetic torque neglected}
Assuming that the magnetic field is not disturbing the Keplerian thin
accretion disk structure, one usually considers the spin-up scenario in
which the angular momentum is transferred from the disk to the NS via accretion from the marginally-stable circular orbit \citep{CookST1994,
ZdunikHG2002}. In this process, an infall of a particle of baryon mass
${\rm d}M_{\rm b}$ and specific orbital angular momentum $l_{\rm ms}$ 
leads to a new quasi-stationary stellar configuration of baryon mass
$M_{\rm b}+{\rm d}M_{\rm b}$ and angular momentum $J+{\rm d}J$, according
to
\begin{equation}
\frac{{\rm d}J}{{\rm d}M_{\rm b}} = x_l l_{\rm ms},
\label{eq:acc_isco}
\end{equation}
where the parameter $x_l$ ($\leq 1$) quantifies our lack of
knowledge of the fraction of angular momentum that is transferred to 
the star by an infalling particle; recent numerical
simulations suggest that the value of $x_l$ is close to unity (see,
e.g., \citealt{BeckwithHK2008} and \citealt{ShafeeMNTGM2008}).

\subsection{Magnetic torque included}
If the magnetic field does affect the accretion flow,
the $B= 0$ model ceases to be correct. To calculate the increase of
the total stellar angular momentum $J$ in this case, one must modify the
Eq.\ (\ref{eq:acc_isco}) to incorporate the fact that the disk now 
terminates at some $r_0>r_{_{\rm ms}}$, depending on the value of
magnetic field. We use the prescription of \citet{KluzniakR2007}, which
gives the following evolution equation:
\begin{equation}
 \frac{{\rm d}J}{{\rm d}M_{\rm b}}=
 l(r_0) -\frac{\mu^2}{9r_0^3 {\dot M}_{\rm b}}
  \left[3-2\left(\frac{r_{\rm cor}}{r_0}\right)^{3/2}\right],
\label{eq:acc_mag}
\end{equation}
where $\mu = BR^3$ is the dipole magnetic moment of an NS (assumed to be
parallel to $J$), $r_{\rm cor}$ is the corotation radius and $\dot{M}_{\rm
b}$ denotes the mean accretion rate; for a detailed description of the
parameters and modifications in obtaining $r_0$, related to the  
relativistic marginally-stable circular orbit, see the appendix and \citet{BejgerFHZ2011}.
\begin{figure}[h]
\resizebox{\columnwidth}{!}
{\includegraphics{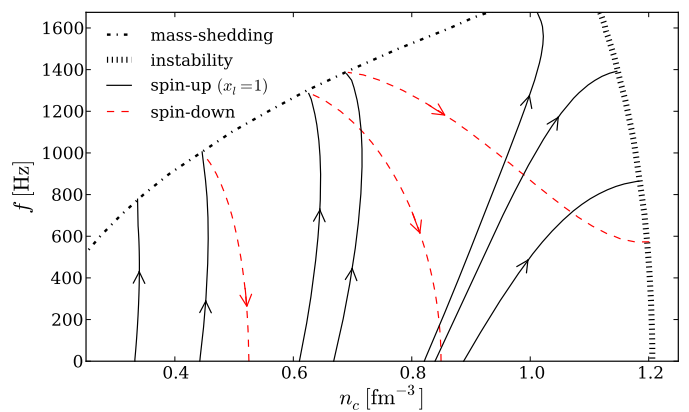}}
\caption{(Color online) Evolution of the central baryon number density
$n_{\rm c}$ in the disk-accretion {\it spin-up} as well as the {\it
spin-down} scenario, shown for the spin frequency $f$--central baryon
density $n_{\rm c}$ plane (DH EOS, $B=0$).}
\label{fig:dh_fnc}
\end{figure}

A complete evolution model of a magnetized NS accreting from a disk 
needs a formula for the accretion-induced surface magnetic field 
decay. Because the theoretical state-of-the-art of this subject is 
quite complicated and far from being complete (see Sect. 4 of 
\citealt {BejgerFHZ2011} for a brief summary), we used the simplest, 
but observationally-motivated decay law proposed by \citeauthor
{ShibazakiMSN1989} (\citeyear{ShibazakiMSN1989}; see also \citealt
{TaamH1986}; \citealt{HeuvelB1995}) and assumed that $B$ decreases 
as a function of accreted mass $\Delta M$ only: $B=B_{\rm i}/(1 + 
\Delta M/{m_B})$, where $m_B=10^{-4}~{\rm M}_\odot$. We also 
tested other available forms of decay law (exponential-like formula 
used recently by \citealt{OslowskiBGB2011} in the context of 
population-synthesis studies, and the quadratic modification of 
\citeauthor{ShibazakiMSN1989} law) and conclude that this choice has 
no qualitative influence on the results.

\section{Results}
\label{sect:results}
To estimate the feasibility of dense-matter phase transitions, we 
examined the compression of matter in the cores of accreting NSs by 
measuring the increase of the central baryon number density $n_{\rm 
c}$ as well as the central baryon chemical potential $\mu_{\rm c} = (p_{\rm 
c} + e_{\rm c})/n_{\rm c}$ ($p_{\rm c}$ and $e_{\rm c}$ are the 
pressure and mass-energy density, respectively). Furthermore, we 
compared the results for {\it spin-up-induced} central compression 
with the isolated NS {\it spin-down} results.

\subsection{Magnetic torque neglected}
For $B=0$ the features of accretion-driven evolution of
$n_{\rm c}$ and $\mu_{\rm c}$ are qualitatively similar for all 
investigated EOSs.
\begin{figure}[h]
\resizebox{\columnwidth}{!}
{\includegraphics{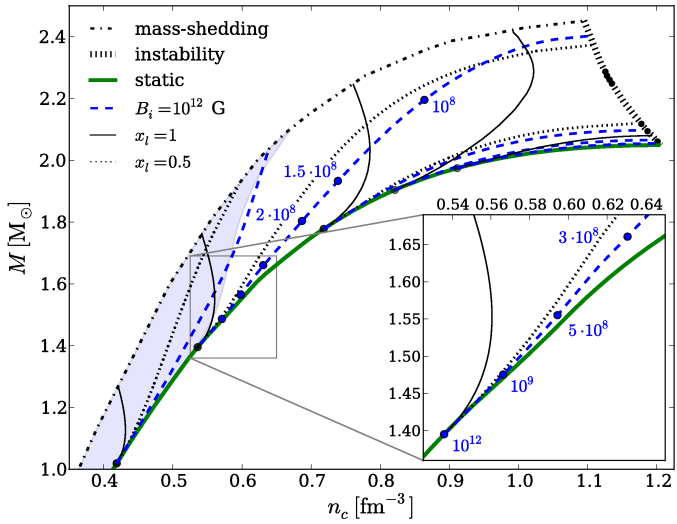}}
\caption{(Color online) Accretion tracks of the gravitational mass--central
baryon number density, $M-n_{\rm c}$, plane. Magnetic field tracks were obtained
for $\dot{M}_{\rm b} = 10^{-9}\ M_\odot/{\rm yr}$. The shaded region contains
configurations for which the stellar equatorial radius $R_{\rm eq}$ is larger
than $r_{\rm ms}$ (DH EOS).}
\label{fig:dh_mnc_xl1}
\end{figure}
We will thus illustrate these features using the DH EOS and begin 
with the comparison of spin-up (with the most efficient angular momentum
transfer $x_l = 1$) and spin-down--induced increase of $n_{\rm c}$
pictured in Fig.~\ref{fig:dh_fnc} in the spin-frequency--central density
plot; the configurations are allowed to span the whole frequency range
from the initially non-rotating stars up to the mass-shedding limit (or
the axi-symmetric perturbation instability limit). Evidently, 
the spin-down allows for a higher $n_{\rm c}$ increase than the $x_l = 1$
accretion spin-up. To investigate this feature, Fig.~\ref
{fig:dh_mnc_xl1} shows the behavior of the gravitational mass $M$ as a
function of $n_{\rm c}$.
\begin{figure}[h]
\resizebox{\columnwidth}{!}
{\includegraphics{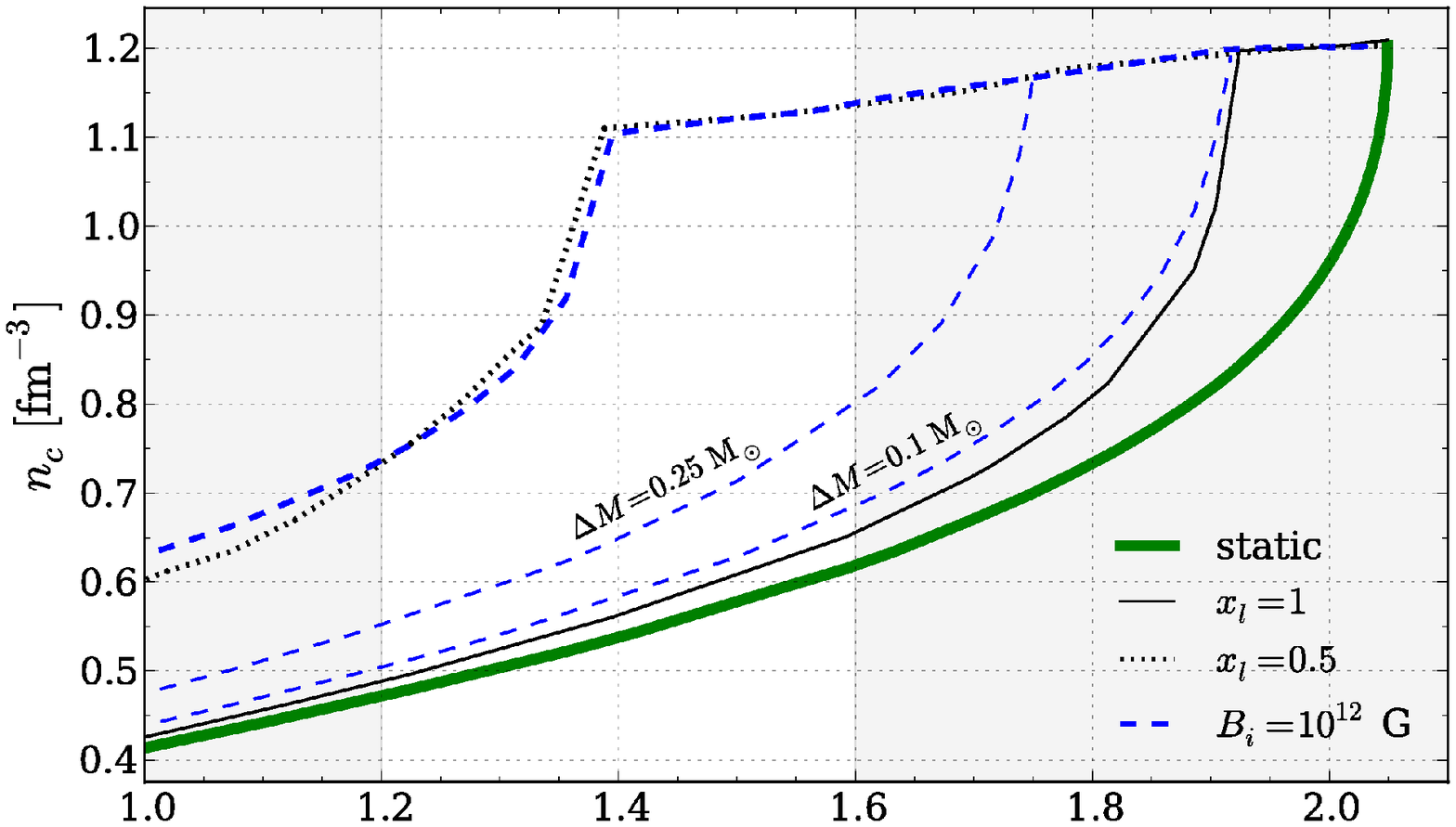}}
\resizebox{\columnwidth}{!}
{\includegraphics{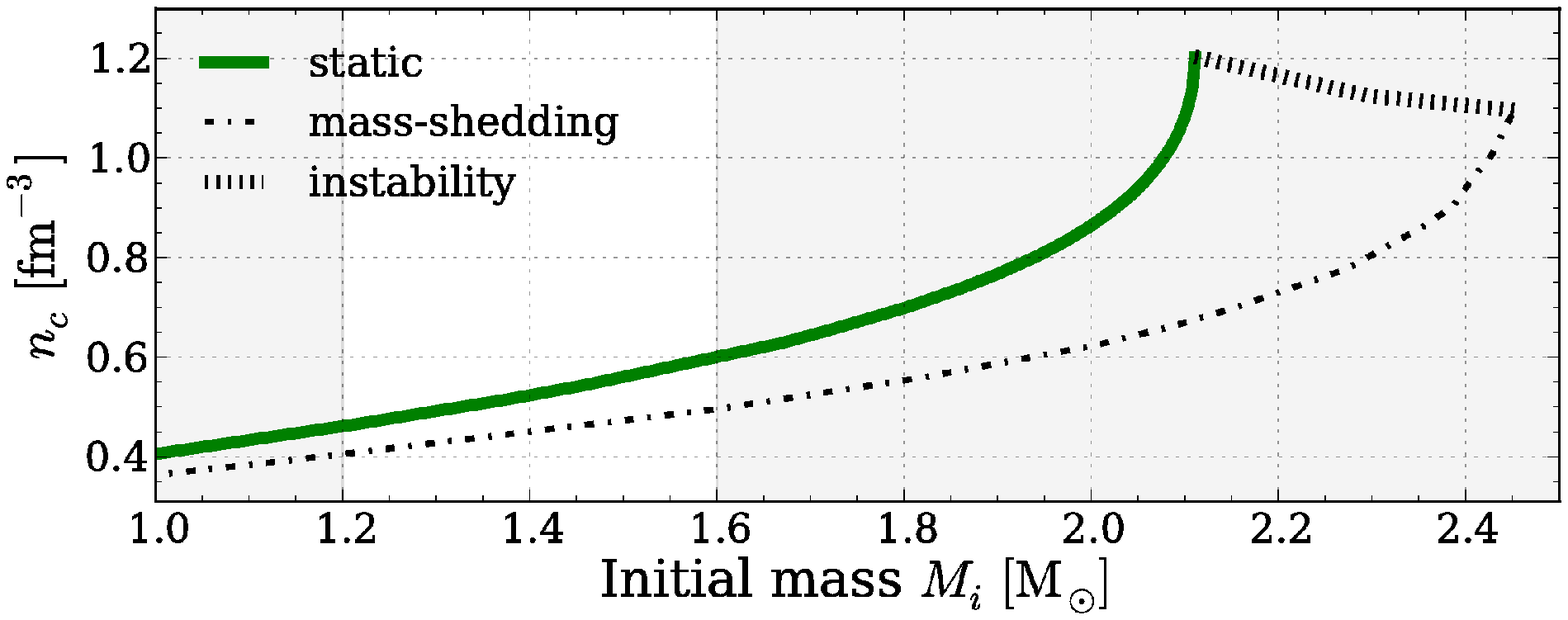}}
\caption{(Color online) Maximum central density $n_{\rm c,f}$ attainable
for a given evolutionary track compared to the initial $n_{\rm c,i}$, both
plotted as functions of the gravitational mass of the {\it initial
non-rotating configuration}, $M_{\rm i}$. {\bf Upper panel}: spin-up
scenario. The reference line $n_{\rm c,i}$ is the static sequence.
{\bf Lower panel}: spin-down scenario. The reference $n_{\rm c,i}$ line
is the mass-shedding limit, to be compared with $n_{\rm c,f}$ of
static configurations (DH EOS).} \label{fig:nc-DH-acc}
\end{figure}
For an astrophysically-motivated range of NS
initial masses, $M_{\rm i}=1.2-1.6~{\rm M}_\odot$, the $x_l = 1$ accretion
never leads to a substantial central compression; moreover, the maximal
central compression is attained for moderate spin periods, followed by a
{\it decrease} of $n_{\rm c}$. On the contrary, lowering the efficiency of
angular momentum transfer to e.g. $x_l = 0.5$ allows for a substantial
$n_{\rm c}$ increase. A detailed summary and comparison between various
types of tracks is included in Fig.~\ref{fig:nc-DH-acc}, where the
initial and final $n_{\rm c}$ is plotted as a function of the {\it
initial} stellar mass $M_{\rm i}$ for both spin-up (upper panel) and
spin-down (lower panel). The nearly horizontal line for $x_l=0.5$ in the upper
panel denotes the instability limit. Based on these data we conclude that to
achieve a significant $n_{\rm c}$ increase in an astrophysically-sound
evolutionary process one should consider 
\begin{itemize}
\item[$\bullet$] an initial mass of the configuration quite close
to the value of the static-star maximum mass in case of $x_l=1$,
\item[$\bullet$] or a value of $x_l$ significantly lower than $1$ (dotted lines in Fig.\ \ref{fig:dh_mnc_xl1} and
\ref{fig:nc-DH-acc}),
\item[$\bullet$] a more efficient mechanism,
such as the {\it spin-down} of an isolated NS - constant $M_{\rm b}$
track leads to a higher {\it increase} of $n_{\rm c}$ than thin-disk
accretion, as shown in Fig.\ \ref{fig:dh_fnc} and the lower panel
of Fig.\ \ref{fig:nc-DH-acc}.
\end{itemize}
Results for the chemical potential $\mu_{\rm c}$ for the DH EOS are plotted
in Fig.~\ref{fig:muc-DH-acc}; results for APR and BM EOSs are presented in
Figs.~\ref{fig:nc-APR-BM-acc} and \ref{fig:muc-APR-BM-acc}.

\subsection{Magnetic torque included}
\begin{figure}[h]
\resizebox{\columnwidth}{!}
{\includegraphics{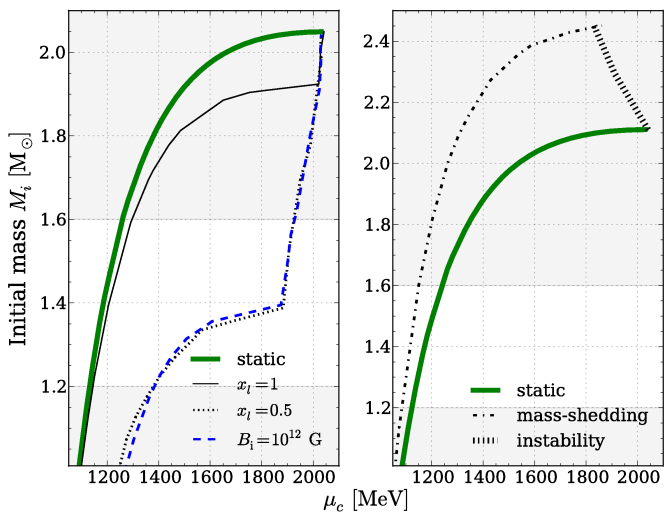}}
\caption{(Color online) Gravitational mass of the initially non-rotating
configuration, $M_{\rm i}$ as a function of the maximum attainable central
baryon chemical potential $\mu_{\rm c,f}$ for a given evolutionary track
compared to the initial $\mu_{\rm c,i}$. Lines are denoted analogously to
Fig.~\ref{fig:nc-DH-acc}. {\bf Left panel}: spin-up scenario. {\bf Right
panel}: spin-down scenario (DH EOS).}
\label{fig:muc-DH-acc}
\end{figure}

\begin{figure}[h]
\resizebox{\columnwidth}{!}
{\includegraphics{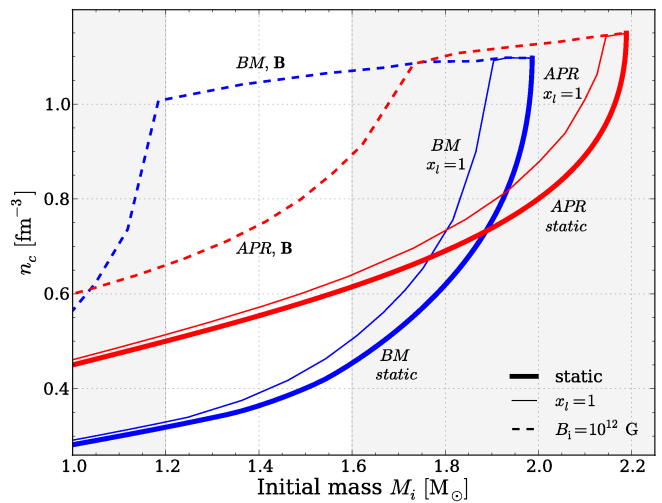}}
\caption{(Color online) Increase of the central baryon number density
$n_{\rm c}$ as a function of the initial gravitational mass $M_{\rm i}$ for
the APR and BM EOSs. Notations as in Fig.~\ref{fig:nc-DH-acc}.}
\label{fig:nc-APR-BM-acc}
\end{figure}

\begin{figure}[h]
\resizebox{\columnwidth}{!}
{\includegraphics{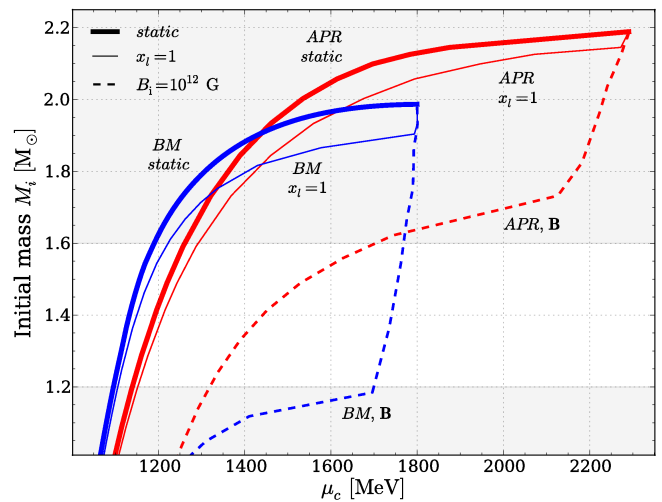}}
\caption{(Color online) initial gravitational mass $M_{\rm i}$ as a function
of intial and maximal central chemical potentials, $\mu_{\rm c}$ for the APR
and BM EOSs. Notations as in Fig.~\ref{fig:muc-DH-acc}.}
\label{fig:muc-APR-BM-acc}
\end{figure}

An inclusion of the magnetic torque acting in the NS-disk system changes the
results significantly, as Fig.~\ref{fig:dh_mnc_xl1} shows. As an
example, we used the ''canonical case'' of the initial magnetic field $B_{\rm
i} = 10^{12}$ G and the average accretion rate $\dot{M}_{\rm b} = 10^{-9}\
M_\odot/{\rm yr}$ (dashed lines; dots denote the decaying magnetic field
for the initial mass $M_{\rm i}=1.4\ M_\odot$). Overall, the
magnetic-torque results resemble those for a reduced efficiency of the
angular momentum transfer, $x_l=0.5$. Including the magnetic field substantially 
increases $n_{\rm c}$ and $\mu_{\rm c}$ for an astrophysically interesting
range of initial masses, $M_{\rm i}=1.2-1.6~{\rm M}_\odot$
(Figs.~\ref {fig:nc-DH-acc} and \ref {fig:muc-DH-acc} for the DH EOS, and
Figs.~\ref {fig:nc-APR-BM-acc} and \ref{fig:muc-APR-BM-acc} for APR and BM
EOSs). The nearly horizontal line segments in the upper panels of Figs.~\ref
{fig:nc-DH-acc} and \ref {fig:nc-APR-BM-acc} correspond to the
axisymmetric instability limit - it is reached by configurations that
were not spun up to the mass-shedding limit; lowering the mass accretion
rate results in an extended instability limit line toward smaller
initial masses. In addition to the maximum attainable compression, the
results for a predefined amount of accreted mass ($\Delta M = 0.1\ M_\odot$, $\Delta M = 0.25\ M_\odot$) are plotted for comparison in Fig.~\ref{fig:nc-DH-acc}.

For the DH EOS and $M_{\rm i}=1.4~{\rm M}_\odot$, accretion of about
$0.25~{\rm M}_\odot$ implies $(n_{\rm c,f}-n_{\rm c,i})/n_{\rm c,i}\approx
0.2$. This should be compared with very small (of a few percent)
compression for $B= 0$ and $x_l=1$ case. Including the magnetic torque is
therefore crucial for accretion-induced compression in the core of an NS
with initial mass $\sim 1.4~\msun$. Note also a fairly strong effect of
the EOS stiffness below approximately three nuclear densities, as well as
its high-density behavior. As seen for example in Fig.~\ref
{fig:nc-APR-BM-acc}, for our fixed magnetic dissipation evolution model, 
a stiff BM EOS with a high-density hyperon softening predicts a fractional increase of $n_{\rm c}$ by about 120\% at $M_{\rm i}=1.4~{\rm M}_\odot$ , while  $\mu_{\rm c}$ increases by $\sim 500~$MeV. The overall
picture obtained for magnetized stars, with $M_{\rm i}=1.2-1.6~{\rm
M}_\odot$, is therefore essentially different from the $B=0$ one.
\section{Conclusions}
\label{sect:conclusion}
We have studied the influence of two thin-disk accretion models - with and
without the involvement of magnetic field - on the evolution of the
central density and chemical potential in accreting NSs. For $B=0$  accretion from the marginally-stable circular orbit leads to a {\it
negligible} central compression (a few per cent) for
astrophysically-relevant initial configurations (slowly-rotating stars
with $M_{\rm i}=1.2-1.6~{\rm M}_\odot$) and maximally-efficient angular
momentum transfer ($x_l=1$ in Eq.~\ref {eq:acc_isco}, a value currently
favored by numerical simulations). Consequently, this mechanism does not
seem viable to trigger a dense-matter phase transition. Substantial
central compression could be obtained for $x_l$ smaller than unity, or in
a different evolutionary process (e.g., spin-down). Including the
magnetic torque changes the outcome profoundly, but for a typical
initial magnetic field, $B_{\rm i}\simeq 10^{12}$ G and accretion rate
$\dot{M}_{\rm b}=10^{-9}\ M_\odot$ one is able to produce massive,
weakly-magnetized ($B\simeq 10^{8}$ G) millisecond pulsars with a sizable
(even as large as $100\%$) central compression, thus probing much higher densities, which in turn may trigger a phase transition. Even such a low final value of $B$ still influences the accretion process and should be accounted for; only lowering it even more, to $\simeq 10^7$ G, essentially allows for the recovery of the $B=0$ results.

Note also a quite impressive dependence of the NS {\it spin-up evolution} on
the very presence of the magnetic-field. The magnetic torque substantially
decreases the spin-up efficiency, as the comparison with the $x_l=1$ and
$x_l = 0.5$ results show - in other words, spinning-up with magnetic field
requires {\it more} accreted mass to reach a desired spin frequency. A
considerable subset of ''magnetic-torque'' tracks ends at the axisymmetric
instability limit (i.e., nearly horizontal lines in
Figs.~\ref{fig:nc-DH-acc} and \ref{fig:nc-APR-BM-acc}), while the vast
majority of the $x_l=1$ tracks can reach the mass-shedding limit. This result
correlates with the EOS stiffness and is most pronounced for the BM EOS,
softened by the existence of hyperons; potentially, it may help in the
understanding of a puzzling non-detection of submillisecond pulsars, as
well as assist in the studies of formation of stellar-mass BHs and their mass
function.

We have restricted ourselves to the EOSs of hadronic matter. 
A phase transition to quark matter softens the high-density EOS, so that
reaching $M_{\rm max}>1.97\ M_\odot$ requires some tuning of both
hadronic and the quark-matter model. An approach based on an effective
model of the QCD of quark matter used recently by \citet{Bonanno2011} 
indicates that the vector repulsion in quark matter should be sufficiently 
strong to reach $M_{\rm max}>1.97\ M_\odot$ (we recall that vector-meson repulsion in hadronic matter is also crucial for our BM
EOS model). In any case, the maximum mass of NSs with quark cores 
(so-called hybrid stars) turns out to be very close to that reached at the 
central density equal to the deconfinement density \citep{Bonanno2011}.

Our results were obtained using three specific EOSs of
dense matter, an effective magnetic torque model in the pulsar--accretion
disk system, and a simplistic description of accretion-induced
magnetic-field dissipation. We believe, however, that in a qualitative
sense these results posses a general validity.
\acknowledgements
This work was partially supported by the Polish MNiSW research grant no. N
N203 512838, LEA Astrophysics Poland-France (Astro-PF) and ESF
Research Networking CompStar programmes. MB acknowledges the Marie Curie
Fellowship within the 7th European Community Framework Programme
(ERG-2007-224793).
\appendix
\section{Calculation of the inner-boundary radius of the accretion disk}
\label{sect:appendix}
To simplify the calculation of the specific orbital angular momentum $l$
at every step of spin-up evolution, we used the \citet{BejgerZH2010}
result: the Keplerian orbital frequency of a particle
in the thin disk at a radius $r_0$ is well-approximated by the
"Schwarzschildian/Newtonian" formula $\sqrt{GM/{r_0}^3}$. This ansatz
yields a surprisingly accurate determination of $l$ for $r_0\simeq r_{\rm
ms}$ for a broad range of stellar masses and spin frequencies up to the
mass-shedding limit.
\begin{figure}[h]
\resizebox{\columnwidth}{!}
{\includegraphics[clip]{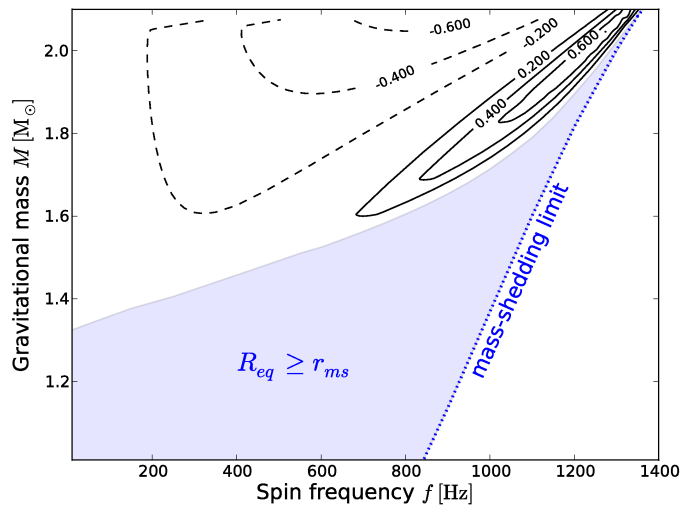}}
\caption{(Color online) Difference (in km) between the real value of 
$r_{\rm ms}$ and the value $r^{\rm (appr.)}_{\rm ms}$ (to be 
compared with average radii of the problem: $R_{\rm eq}, r_{\rm ms} 
\sim 10$ km) obtained by means of an approximation of \citet
{BejgerZH2010, BejgerFHZ2011}, plotted as iso-contours on the $M$--
$f$ plane (DH EOS).} 
\label{fig:rms-approx}
\end{figure}
Here we show that this approximate approach allows for a
reasonably accurate determination of the $r_{\rm ms}$ as well. It corresponds to the ${\rm d}l/{\rm d}r=0$ condition; the inner-edge of
an accretion disk $r_0$ is a solution of
\begin{equation}
\label{bc}
\frac{1}{2} f_{\rm ms}(r_0) = \left(\frac{r_m}{r_0}\right)^{7/2}
\!\! \left(\sqrt{\frac{r_{\rm cor}^3}{r_0^3}}-1\right) =
{\xi^{7/2} \omega^{-10/3}\,\,(1-\omega)},
\end{equation}
where $r_{\rm cor}$ is the corotation radius, $r_m = (GM)^{-1/7}\dot
M^{-2/7} \mu^{4/7}$ is the magnetospheric radius and $\xi = r_m/r_{\rm cor}$.
Function $f_{\rm ms}$ introduced by \citeauthor{BejgerFHZ2011} (\citeyear{BejgerFHZ2011}; reducing to
$f_{\rm ms}\equiv 1$ in the Eq.~17 of \citealt{KluzniakR2007}) ensures
a proper behavior of the torque near $r_{\rm ms}$ 
\begin{equation}
\label{eq:frel}
f_{\rm ms}=\frac{1-\alpha/{\bar r}^{3/2}}{(1-v^2/c^2)^{3/2}
\sqrt{1-1/{\bar r}}}\left(\frac{{\bar r}-2}{{\bar r}-1}-2
\frac{v^2}{c^2}+\frac{3\alpha}{{\bar r}^{3/2}-\alpha}\right),
\end{equation}
with $\alpha=Jc/(\sqrt{2}GM^2)$, ${\bar r}={r_0}/{r_{s}}$ and $r_s = 2GM/c^2$
denoting the Schwarzschild radius. Marginally stable circular orbit radius
$r_{\rm ms}$ is then a solution of equation $f_{\rm ms}(r_{\rm ms})=0$.
Fig.~\ref{fig:rms-approx} shows, using the example of the DH EOS (other EOSs
we tested yield similar results) the comparison of the $r_{\rm ms}$
calculated {\it exactly} from integrals of motion with the approximate
value $r^{\rm (appr.)}_{\rm ms}$, obtained by means of Eq.~(\ref
{eq:frel}). For almost all configurations the difference is less than
0.6 km (for $r_{\rm ms}<R$ we adopt $r_{\rm ms} = R$), and the regions
where $|r_{\rm ms}-r^{\rm (appr.)}_{\rm ms}|>0.6$~km are small and
well-confined near high-mass and mass-shedding, submillisecond rotation.

\end{document}